\documentclass[preprint,review,12pt]{elsarticle}
\usepackage{amssymb}
\usepackage{amsmath}
\usepackage{lineno}
\usepackage{graphicx}
\usepackage{url}
\usepackage{comment}
\usepackage{xcolor}
\usepackage[output-decimal-marker={.}]{siunitx}
\DeclareSIUnit\barn{b}
\DeclareSIUnit\ton{t}
\DeclareSIUnit\angstrom{\text{Å}}
\DeclareSIUnit\bar{bar}
\usepackage{adjustbox}
\usepackage{float}
\usepackage{subfigure}
\usepackage{caption}
\usepackage{subcaption}
\usepackage{amsmath} 
\usepackage{amsfonts}
\usepackage{fancyhdr}
\usepackage{csquotes}
\usepackage{graphicx}
\usepackage{tikz}
\usepackage{listings}
\usepackage{wrapfig}

\makeatletter
\def\ps@pprintTitle{%
  \let\@oddhead\@empty
  \let\@evenhead\@empty
  \let\@oddfoot\@empty
  \let\@evenfoot\@oddfoot
}
\makeatother

\begin{document}

\begin{frontmatter}

\title{Capturing methane in a barn environment: \\ the CH$_4$ Livestock Emission (CH4rLiE) project}

\author[a,b]{Francesco Alessandro Angiulli}
\author[b]{Chiara Aimè~\fnref{1}}
\author[c,d]{Maria Cristina Arena}
\author[e,f]{Davide Biagini}
\author[b]{Alessandro Braghieri}
\author[a,b]{Matteo Brunoldi}
\author[a,b]{Simone Calzaferri}
\author[e,f]{Elio Dinuccio}
\author[c,b]{Daniele Dondi}
\author[f]{Linda Finco}
\author[d]{Roberto Guida}
\author[a]{Nithish Kumar Kameswaran}
\author[d]{Beatrice Mandelli}
\author[a,b]{Paolo Montagna}
\author[a,b]{Cristina Riccardi}
\author[b]{Paola Salvini}
\author[b]{Alessandro Tamigio}
\author[a,b]{Ilaria Vai~\corref{cor1}} \ead{ilaria.vai@unipv.it}
\author[c]{Dhanalakshmi Vadivel}
\author[f]{Riccardo Verna}
\author[a,b]{Paolo Vitulo}

\fntext[1]{Now at Dipartimento di Fisica, Università di Pisa and INFN Sezione di Pisa, Largo Bruno Pontecorvo~3, 56127 Pisa, Italy}
\cortext[cor1]{Corresponding author}

\affiliation[a]{organization={Dipartimento di Fisica, Università di Pavia},
            addressline={Via Bassi~6}, 
            city={Pavia},
            postcode={27100}, 
            state={},
            country={Italy}}
\affiliation[b]{organization={Istituto Nazionale di Fisica Nucleare, Sezione di Pavia},  addressline={Via Bassi~6}, 
            city={Pavia},
            postcode={27100}, 
            state={},
            country={Italy}}
\affiliation[c]{organization={Dipartimento di Chimica, Università di Pavia},
            addressline={Via Torquato Taramelli~12}, 
            city={Pavia},
            postcode={27100}, 
            state={},
            country={Italy}}
\affiliation[d]{organization={CERN},
            addressline={Esplanade des Particules~1}, 
            city={Geneva},
            postcode={1211}, 
            state={},
            country={Switzerland}}
\affiliation[e]{organization={Dipartimento di Scienze Agrarie, Forestali e Alimentari, Università di Torino},
            addressline={Largo Paolo Braccini~2}, 
            city={Grugliasco (TO)},
            postcode={10095}, 
            state={},
            country={Italy}}
\affiliation[f]{organization={Istituto Nazionale di Fisica Nucleare, Sezione di Torino},
            addressline={Via Pietro Giuria~1}, 
            city={Torino},
            postcode={10125}, 
            state={},
            country={Italy}}

\begin{abstract}

The CH$_4$ Livestock Emission (CH4rLiE) project explores the development of a prototype system for capturing methane emissions in barn environments, offering an alternative approach to mitigating greenhouse gas emissions from livestock farming. Methane (CH$_4$), with a global warming potential significantly higher than CO$_2$ (GWP$_{100}$ = 27), accounts for $\sim$23\% of anthropogenic climate impact. In 2021, The Assessment Report~6 of Intergovernmental Panel on Climate Change quantified CH$_4$ livestock emissions in 123 Mt/yr, which, together with substantial N$_2$O and CO$_2$ emissions, contributed with a 12\% to global emissions.
Unlike strategies focused on altering animal feed, CH4rLiE investigates post-emission capture using porous materials, such as zeolites, to adsorb methane from barn air. The project draws on CERN’s experience with gas recovery systems for particle detectors, adapting similar technologies to agricultural settings. Preliminary estimates, based on measured CH$_4$ concentrations ($\sim$20 mg/m$^3$) and partial air filtration in a 250-animal barn, suggest a low but detectable recovery potential, subject to validation through simulation and in-situ testing.
Prototype development considers the potential for energy-efficient operation - possibly through pressure swing regeneration - and compatibility with existing ventilation infrastructure, though these aspects remain under evaluation. If methane concentrations in barns prove too diluted, the system may be better suited for environments with higher gas levels, such as pigsties or landfills. NH$_3$ capture for fertilizer production is planned as a future enhancement.
CH4rLiE aims to assess the feasibility of emission recovery in livestock settings without affecting animal welfare, contributing to sustainable farming practices, resource efficiency, and circular bioeconomy goals.

\end{abstract}

\begin{keyword}
methane \sep capture \sep livestock emissions \sep global warming

\end{keyword}

\end{frontmatter}

\section{Introduction}
\label{sec:intro}
“The Global Methane Budget 2000-2020”~\cite{saunois} published in 2025 quantifies the relevant role played by methane (CH$_4$) in global warming. Nowadays, CH$_4$ is the second most important human-influenced greenhouse gas in terms of climate forcing, just after carbon dioxide (CO$_2$). In addition, its atmospheric emissions and concentration continue to increase due to contributions from both natural and anthropogenic sources. Notably, methane current growth is dominated by emissions from biological sources, such as wetlands, waste and agriculture~\cite{nisbet2}. Global emissions from agriculture and waste for the period 2010–2019 are estimated to be 211~[195–231]~TgCH$_4$yr$^{-1}$, representing 60\% of total direct anthropogenic emissions. For the period 2010–2019, emissions from enteric fermentation and manure management were estimated in 112~[107–118]~TgCH$_4$yr$^{-1}$, about one-third of total global anthropogenic emissions~\cite{saunois}.

In this framework, the CH$_4$ production happens mainly during the anaerobic digestion of ruminants (87-90\% expelled from the mouth~\cite{broucek}) and during the management of manure, which can release CH$_4$ by anaerobic fermentation, if rich in water. Otherwise, the aerobic fermentation of dry manure releases nitrous oxide (N$_2$O), which has a global warming potential over 100 years (GWP$_{100}$) much higher than CH$_4$ (265 vs 27)~\cite{ggp}. The production of this last gas can anyway be avoided by disposal of manure in anaerobic digesters for the production of biogas. 
Addressing CH$_4$ emissions from enteric fermentation is therefore of primary importance. CH$_4$ release can be mitigated either directly, for instance through the use of feed additives that inhibit methanogenesis, or by removing the gas from the air after it has already been emitted~\cite{nisbet2}. Removal processes, however, require energy, and their implementation must result in a net positive impact on global warming mitigation to be considered viable. Additionally, any proposed technology must also ensure that the well-being of the animals is not compromised.
A solution must therefore be sought in recovering the CH$_4$ component once already emitted into the enclosed environments, such as stables, in which animals stay for long periods.
The so-called Direct Air Capture (DAC)~\cite{bisotti} techniques refer to a set of technological solutions designed to remove greenhouse gases already released into the atmosphere. They are being extensively tested for CO$_2$ removal, but could represent an interesting approach also for CH$_4$. 

The methodologies employed in DAC techniques include:
a) Chemical Capture: air is passed through a chemical medium that reacts with the target gas (CO$_2$ or CH$_4$). Once the gas has been captured, it is separated from the
medium and can be stored or used. It is effective for very low CH$_4$ concentrations, but separating and purifying the captured CH$_4$ is energy intensive.
b) Physical Capture: air is passed over porous materials such as zeolites or activated carbon, which adsorb the CH$_4$. After an adsorption cycle, the CH$_4$ is released
for purification or storage. The process is less expensive than a), but the efficiency depends on the ability of the material to capture and hold the CH$_4$.
c) Electrochemical Capture: electrochemical cells are used to capture CH$_4$, which is oxidized to CO$_2$ in a process that can produce energy or other useful byproducts.
It can be used in smaller systems, such as for confined air purification, but the technology is still at a very early stage of research.
A few projects~\cite{FIS03_18, FIS03_19, FIS03_20} successfully applied DAC techniques to CO$_2$ capture and are considering a future application to CH$_4$ as well as to other greenhous gases (GHGs).

In this paper, we present the CH$_4$ Livestock Emission (CH4rLiE) project, which aims to develop a prototype system for capturing methane emissions in a barn environment by means of adsorption onto porous materials such as zeolites. This technique, already widely used in industrial applications, offers a significant advantage due to its potential for rapid scalability, which is an important factor given the urgency of decarbonization~\cite{nisbet1}. In parallel, the project explores novel porous materials to further enhance methane adsorption capacity.

\section{The CH$_4$ Livestock Emission project}\label{sec:ch4rlie}

Compared with CO$_2$, CH$_4$ recovery with DAC techniques from a barn environment presents a major challenge related to the fact that the CH$_4$ concentration in this atmosphere is quite low (preliminary measures indicate an order of 20 mg/m$^3$). In this context, ongoing work on the recovery of pollutant gases used in mixtures for particle detectors exploited in high-energy physics experiments becomes relevant. In recent years, CERN~\cite{cern} has been actively working to reduce pollutant emissions from its experimental activities~\cite{guida1}: 92\% of these emissions were related to experimental activity, particularly due to the use of high-GWP gases in particle detectors. The most commonly used components, in this case, are C$_2$H$_2$F$_4$ (GWP$_{100}$ = 1430), SF$_6$ (GWP$_{100}$ = 24300) and CF$_4$ (GWP$_{100}$ = 7380)~\cite{supplementary}. Along with extensive R\&D aimed at finding alternative components, a mitigation strategy followed with extreme success was the implementation of ``gas recuperation systems''. Through these systems, the component of interest in the mixture is extracted, stored, and, when necessary, reused. The typical high level of complexity of such a plant is repaid by the excellent results that can be obtained: recent outcomes show that it is possible to recover up to 80\% of C$_2$H$_2$F$_4$ present in the mixture of Resistive Plate Chambers~\cite{santonico}, gaseous detectors widely used in high energy physics experiments.

CH4rLiE aims therefore at adapting a gas recuperation system~\cite{guida2} developed in the framework of the Large Hadron Collider (LHC)~\cite{lhc} experiments at CERN to capture CH$_4$ emitted in a barn environment. The major challenge is then to retrofit the system to the barn environment, particularly moving from gas concentrations to be recovered on the order of \% to a few~ppm.

\subsection{Characterization of the barn atmosphere}\label{subsec:simulation}

Complete characterization of the barn atmosphere is essential to verify the feasibility of installing the methane capture prototype and to quantify its impact in terms of the amount of methane extracted. In the context of the CH4rLiE project, this characterization is done by direct measurements in the barn, which are then compared to results from fluid dynamic simulations.

\subsubsection{Direct seasonal monitoring} 

Direct measurements in the barn environment were carried out in 2023 and 2024 to analyze variations in air quality composition throughout the day and across different seasons. The work, led by the research group of the Department of Agricultural, Forest and Food Sciences of the University of Torino, enabled the assessment of environmental parameters. 

\begin{figure}
    \centering
    \includegraphics[width=0.7\linewidth]{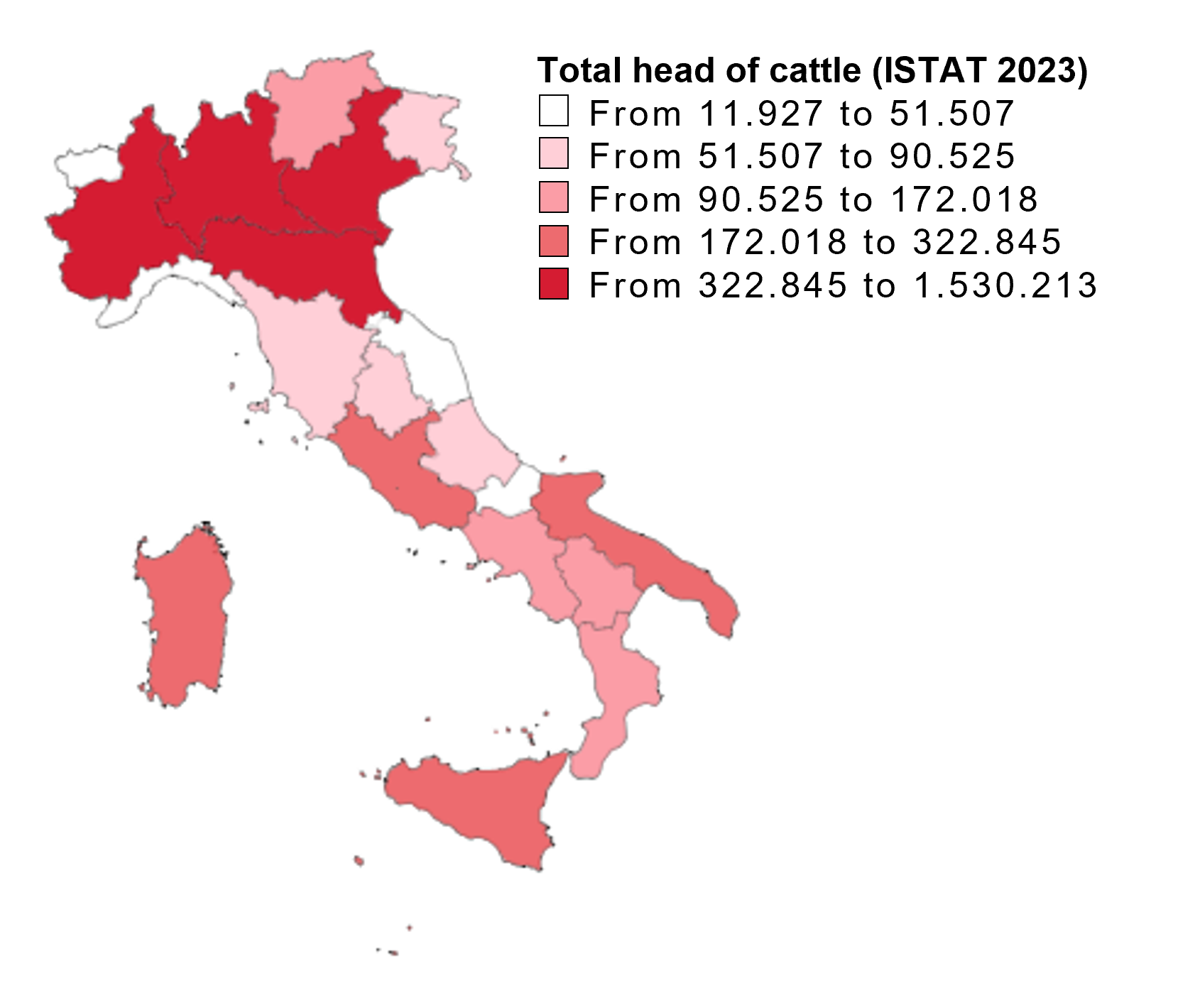}
    \caption{Head of cattle by region in Italy.}
    \label{fig:cattle_map}
\end{figure}

\begin{figure}
    \centering
    \includegraphics[width=0.9\linewidth]{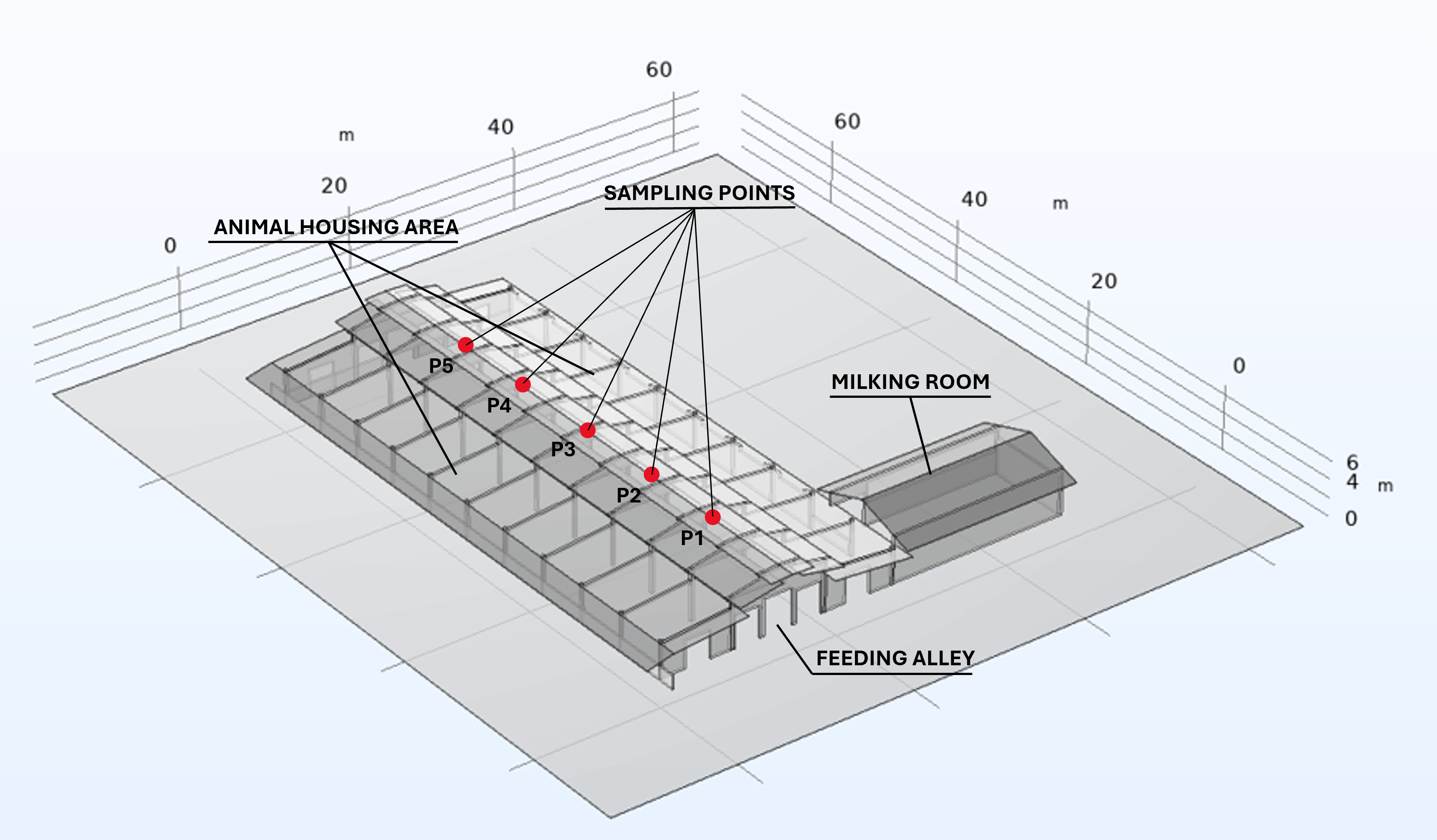}
    \caption{Simplified layout of the barn in Volvera chosen for the measurement campaigns.}
    \label{fig:barn}
\end{figure}

The barn chosen for the measurements is a dairy cattle facility located in Volvera, near Torino, in Northen Italy, an area where intensive dairy cattle production is largely concentrated, as shown in Fig.~\ref{fig:cattle_map}. The barn features an open housing system, allowing for natural ventilation, and hosts over 300 cows. Manure is automatically removed four times per day, meaning that the methane present in the barn atmosphere is primarily emitted by the animals themselves. The layout of the barn is illustrated in Fig.~\ref{fig:barn}.

The analysis of CH$_4$ emissions is conducted using the Quartz-Enhanced Photoacoustic Spectroscopy (QEPAS) technique, which enables the selective measurement of a target gas. The QEPAS analyzer measures the vibrational
relaxation rate of gas species by employing quartz tuning forks (QTFs) as sound detectors. Owing to its high-quality factor and narrow resonance bandwidth, the QTF allows for the detection of weak photoacoustic signals
with improved selectivity and strong resistance to ambient acoustic noise. As reported in~\cite{menduni}, QEPAS-based systems can achieve detection limits as low as 30 ppb for CH$_4$, providing the sensitivity required to measure methane concentrations in the reference barn. Additionally, they feature low power consumption and minimal cross-sensitivity to interfering gases.
The QEPAS detector employed in these measurements is coupled with a multi-point sampler and sequentially draws air samples from five different locations along the central corridor of the barn, called $Pi$ (with $i = 1,2,3,4,5$). These measurement points are spaced 10 meters apart and positioned at a height of 5 meters. For each point, the analyzer determines the concentration of the target gas and provides a value expressed in ppm. After completing the full sampling cycle, the system restarts from the first point. Four measurement campaigns are carried out over the course of a year, covering seasonal variability. Each campaign consist of 24 hours of continuous data collection. The goals of the measurements are the evaluation of gaseous emissions and the monitoring of CH$_{4}$ concentration. These results serve as a fundamental reference for current and future studies, providing essential data to improve emission mitigation systems.

To estimate daily methane emissions, the mass balance approach based on CO$_2$ measurements is used~\cite{pedersen}. In addition to enabling CH$_4$ flux estimation, this approach offers a more comprehensive understanding of carbon gas dynamics in animal housing systems, potentially
accounting for both emission outputs and carbon immobilization processes~\cite{massbalance1, massbalance2}.

\begin{figure}[!h]
    \centering
    \includegraphics[width=1\linewidth]{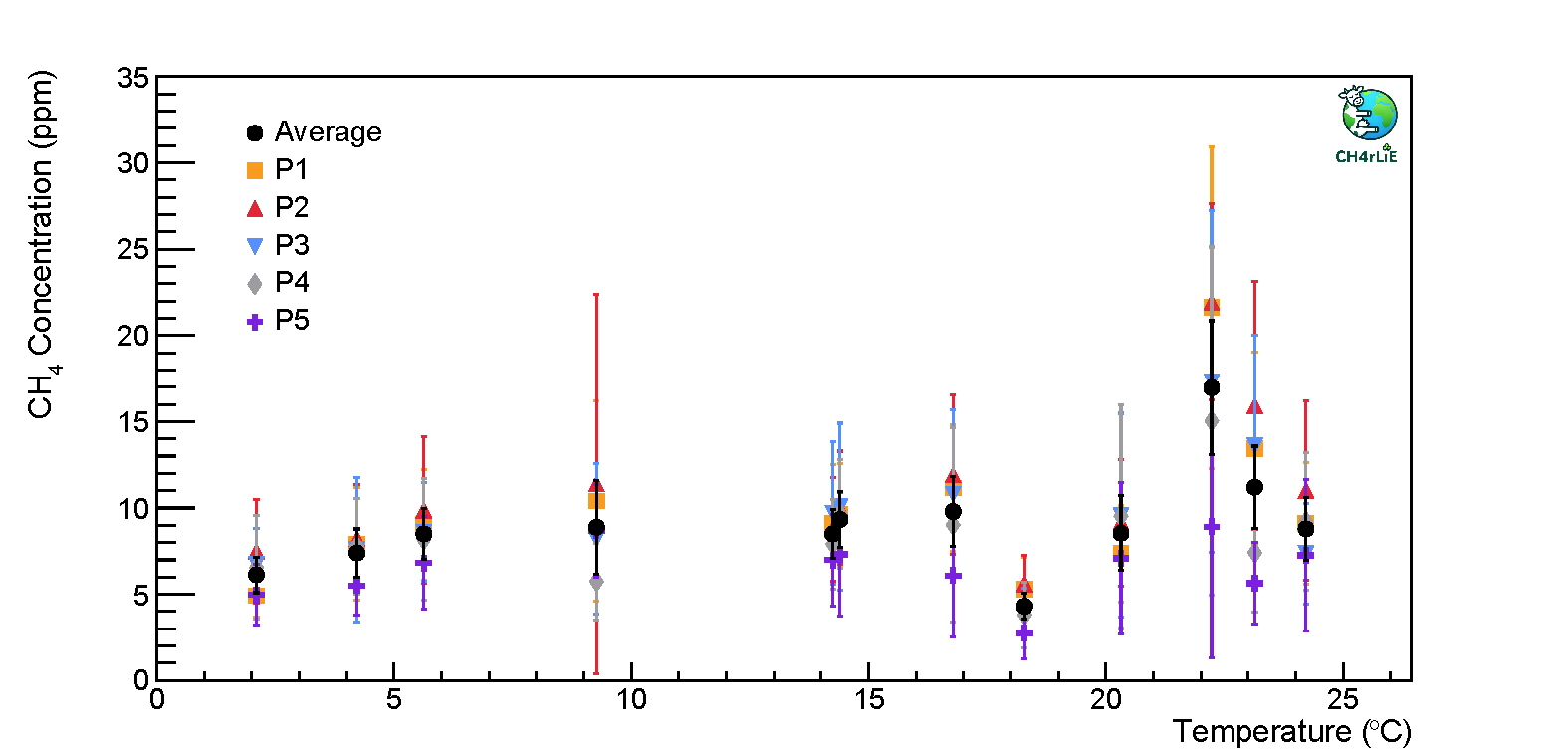}
    \caption{Average concentration of CH$_4$ in 5 measurement points along the barn as a function of the mean temperature recorded inside the barn. The error bars are the standard deviation of each sample.}
    \label{fig:concCH4_vs_temp}
\end{figure}

\begin{figure}[!h]
    \centering
    \includegraphics[width=1\linewidth]{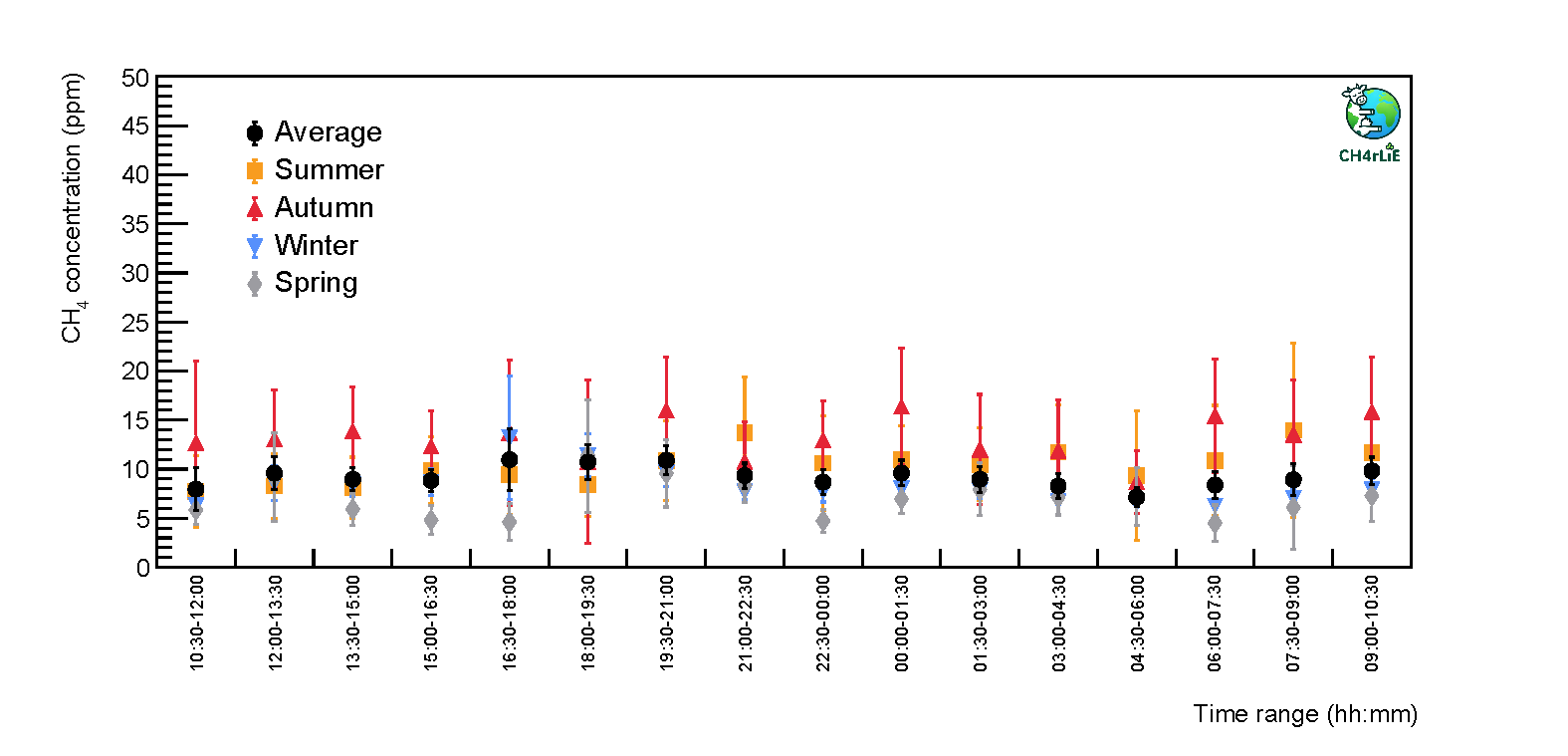}
    \caption{Concentration of CH$_4$ measured over a 24-hour period, shown separately for each season and for the overall average. The error bars are the standard deviation of each sample.}
    \label{fig:concCH4_vs_time}
\end{figure}

Fig.~\ref{fig:concCH4_vs_temp} shows an example of the variation in methane concentration at different measurement points as a function of the average temperature recorded inside the barn. Methane levels range from a few ppm up to approximately 20 ppm. These relatively low concentrations are primarily due to the distance between the animals and the sampling points, which are positioned 5 meters above the ground. Moreover, within the uncertainties (computed as the standard deviation of each sample) the average CH$_4$ concentration does not exhibit significant variation. Notably, higher methane levels are generally observed at points 1 and 2, which are located closest to the barn entrance. This may be attributed to their proximity to the milking area, where the cattle voluntarily gather twice per day. These observations suggest that the methane capture prototype should be positioned near the milking room and, if feasible, installed closer to the animals. Fig.~\ref{fig:concCH4_vs_time} shows the average concentration of CH$_4$ measured over a 24-hour period, presented separately for each season and for the overall average. The data represent the mean values across the five different sampling points. As shown, average CH$_4$ concentrations do not exhibit significant diurnal variation. Measurements shown in Fig.~\ref{fig:concCH4_vs_temp} and~~\ref{fig:concCH4_vs_time} will be compared with data to be collected by more precise monitoring stations, which are scheduled for installation in the coming months.

\subsubsection{Simulations of methane diffusion}

Numerical simulations were performed to analyze methane diffusion within a dairy barn, based on a realistic geometric reconstruction. To reduce computational complexity, a progressive modeling approach was adopted: starting with a simplified 2D longitudinal cross-section of the barn and gradually introducing structural and physical complexity.

\begin{figure}
\centering
\includegraphics[width=0.5\textwidth]{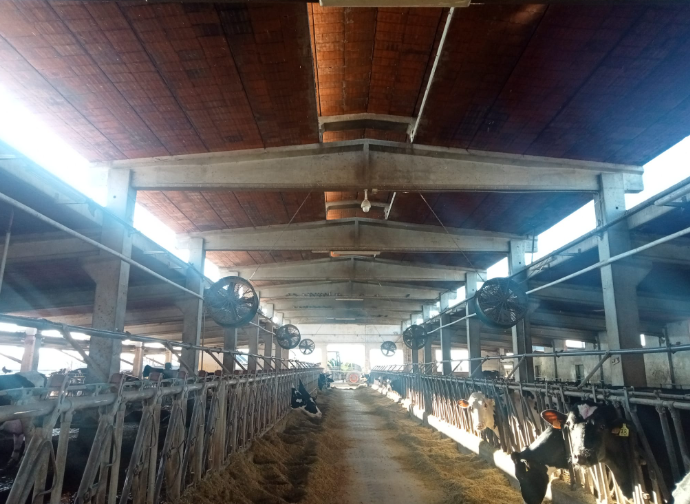}
\caption{Inside picture of the real barn}
\label{fig:real_barn_pictures}
\end{figure}

The initial model represents a vertical section along the barn’s central corridor, shown in Fig.~\ref{fig:real_barn_pictures}, with pens on both sides and a connected milking parlor. In high-yield dairy farms, a single cow can emit up to 120 kg CH$_4$ per year ($\sim$330~g/day)~\cite{Cow_CH4_production_data}. For simulation purposes, cows are modeled as isotropic point sources emitting methane uniformly over time.

\begin{figure}
    \centering
    \includegraphics[width=1\linewidth]{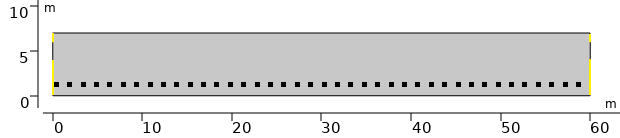}
    \caption{2-D geometry representing the longitudinal section of the barn. Yellow lines represent the open boundaries: two doors and two opening under the roof}
    \label{fig:Geometria2D}
\end{figure}

The 2D model includes 40 such point sources (distributed at 1.2~m height) emitting at a rate of \SI{2.38e-5}{mol\per\second}. 

Methane transport was modeled using the Transport of Diluted Species (TDS) interface in COMSOL Multiphysics~\cite{Comsol_TDS, Comsol_homepage}. The simulation assumes low methane concentrations, allowing use of Fick’s Law for diffusive flux:

\begin{equation}
\vec J = -D \vec \nabla c,
\end{equation}
where $\vec J$ [\unit{mol\per\square\metre\second}] is the diffusive flux vector, $D = \SI{2.1e-5}{\square\metre\per\second}$~\cite{methane_diffusion_coeff} is the diffusion coefficient and $c$ [\unit{mol\per\cubic\metre}] is the methane concentration~\cite{Comsol_TDS}.

Combined with Fick’s law, the general species transport equation includes both diffusion and convection:

\begin{equation}
\frac{\partial c}{\partial t } + \vec \nabla \cdot \vec J + \vec u \cdot \vec \nabla c = R,
\end{equation}
where $\vec u\cdot  \vec\nabla c$ represents convective transport due to fluid motion, with velocity $\vec u$ [\unit{\metre\per\second}], and $R$ [\unit{mol \per\cubic\metre\second}] is the reaction term, which accounts for any sources or consumption of the chemical species~\cite{Comsol_TDS}.

In this initial model, no airflow ($\vec u$ = 0) or chemical reactions ($R$ = 0) are included. Boundary conditions are set as open boundaries at the barn’s doors and roof vents (see Fig.~\ref{fig:Geometria2D}), and initial methane concentration is uniform at the atmospheric level of 2 ppm (\SI{7.8e-5}{mol\per\cubic\metre})~\cite{AR6_IPCC}.

\begin{figure}[h]
    \centering
    \includegraphics[width=1\linewidth]{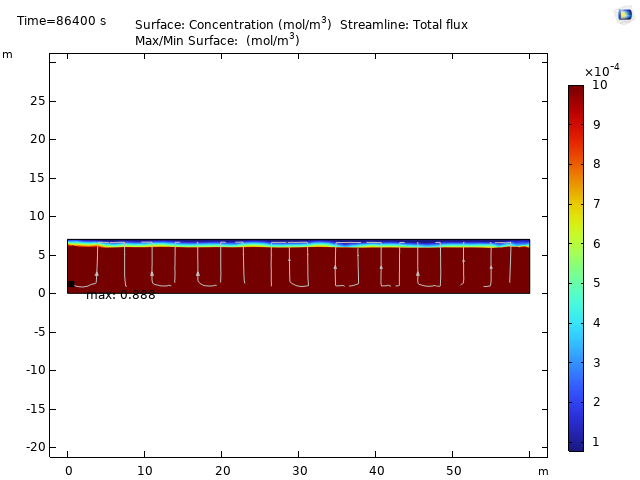}
    \caption{Methane concentration field obtained considering only diffusion. Emission rate of each point source set to \SI{2.38e-5}{\mole\per\metre\per\second}}
    \label{fig:risultati_soloTDS}
\end{figure}

A 24-hour time-dependent simulation was performed to assess how methane diffuses from the sources into the barn atmosphere in the absence of ventilation. This configuration allows for an evaluation of the time scale over which methane concentration departs from the background level. The results, shown in Fig.~\ref{fig:risultati_soloTDS}, indicate significant accumulation near the emission sources and lower concentrations toward the roof, consistent with diffusion-dominated transport.

Finally, in this study, we chose not to include variable atmospheric conditions - such as changes in temperature, humidity, or wind - in our simulations. This choice is justified by the nature of the experimental data, which consist of methane concentration measurements averaged over multiple days throughout the year. Since these measurements integrate a wide range of environmental conditions, simulating specific atmospheric scenarios would not improve the validity of the comparison and could instead introduce unnecessary complexity.

The only environmental detail we plan to incorporate is artificial ventilation, which is commonly used in summer to enhance air circulation in barns. Given its known and regular implementation, ventilation represents a meaningful addition to the model that can influence methane dispersion without conflicting with the time-averaged nature of the experimental data. The implementation of artificial ventilation is currently ongoing in the more realistic 3D geometry of the barn model.

\subsection{Research on adsorption material}\label{subsec:chem}

Zeolites are attractive materials for CH$_4$ capture due to methane’s non-polarity and weak interaction with most surfaces, which limits the effectiveness of conventional liquid solvents~\cite{kim}. Among solid adsorbents, copper-doped zeolites have shown high efficiency in oxidizing CH$_4$ to CO$_2$ under atmospheric conditions and low concentrations, operating at relatively low temperatures (200–300°C) with extended stability and low material cost~\cite{brennis}.

In agricultural contexts such as livestock farming, CH$_4$ constitutes 15–30\% of gaseous emissions, typically alongside CO$_2$, H$_2$, and N$_2$O~\cite{popa}. Reducing methane emissions can generate up to \$4300 per tonne in societal benefits~\cite{key_findings}. However, challenges remain in separating CH$_4$ from these mixtures and minimizing the energy required for its selective capture.

This project focuses on evaluating materials capable of selectively adsorbing CH$_4$ in the presence of common barn gases. Initial screening will use commercial molecular sieves, followed by investigation of novel porous materials- particularly zeolites and pyrolytic carbon - at the Laboratory of Radiation Chemistry and EPR Spectroscopy at the University of Pavia~\cite{kim,yang,nexus2025}.

Carbon-based adsorbents are hydrophobic, which reduces water interference, while zeolites, though more polar, typically offer higher methane uptake. Zeolites also benefit from high thermal and chemical stability and are synthesized using low-cost precursors under hydrothermal conditions~\cite{khaleque}. Their adsorption properties can be tuned through framework composition (e.g., Si/Al ratio), heteroatom substitution (e.g., Ga, Ge), and structural modifications.

Second-generation zeolites such as gallium-germanium frameworks remain underexplored, but show potential for enhanced CH$_4$ selectivity and capacity~\cite{nexus2025}. Further performance improvements may be achieved through techniques like metal ion doping, pore size tuning~\cite{ref26}, and integration with activated carbon~\cite{ref28} or metal-organic frameworks (MOFs)~\cite{ref29}.

In parallel, the project is examining pyrolytic carbon materials derived from sustainable sources such as lignin and agricultural biomass. These are produced under vacuum at temperatures up to 1100°C, with properties influenced by feedstock composition, inorganic content, and thermal processing parameters.

Adsorption performance are being assessed at ambient pressure using methane-containing gas mixtures. Materials are exposed to the gas in closed systems, and uptake is quantified via quadrupole gas analysis. This setup enables a direct comparison of adsorption efficiency across commercial and synthesized materials under realistic conditions.

\subsection{Preparation, testing and deployment of a prototype for CH$_4$ capture}\label{subsec:prototype}

\begin{figure}[!h]
    \centering
    \includegraphics[width=0.9\linewidth]{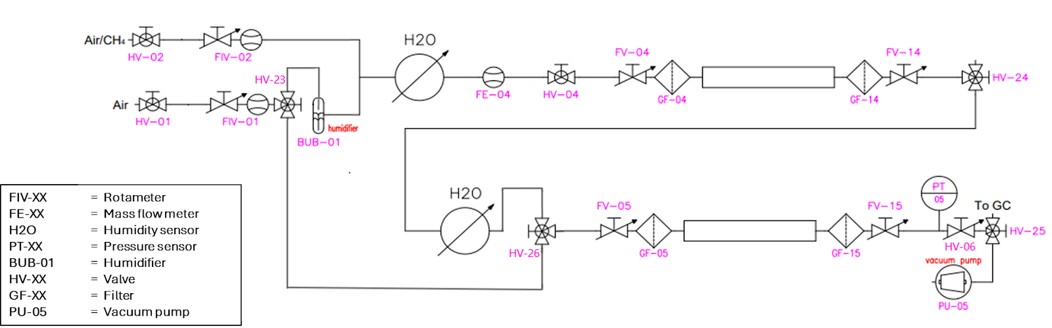}
    \caption{Setup used to test CH$_4$ capture and filtration of pollutant gases. Details in the text.}
    \label{fig4}
\end{figure}

The main target of the project is the design, assembly and deployment of the CH$_4$ capture prototype. The prototype, sketched in Fig.~\ref{fig4}, will be designed on the basis of an already existing CF$_4$ recovery plant~\cite{guida2}\cite{guida3}\cite{mandelli}.

In the testing phase, a gas mixer is used to prepare the desired test gas mixture with air, CH$_4$ and other pollutants which are present in the barn (CO$_2$, N$_2$O, NH$_3$ and humidity). The quality of the gas mixture can be checked using a dedicated extraction line for the analysis. A first column is used to remove humidity from the gas mixture, down to a concentration of a few ppm. A second column is then filled with the adsorbing material selected for CH$_4$ capture. Finally, an extraction line can be delivered to a gas chromatograph for analysis.
When the system is operated, the gas is flushed inside the columns: target molecules (e.g. water in the first column) remain trapped in the adsorber material by the effect of Van Der Waals interaction of gas molecules with the porous material. When the material has reached full adsorption capacity, it loses its adsorption power and has to be regenerated. This operation can be performed either by extracting the trapped molecules with a vacuum pump, or, if they are too strongly bound to the material, by heating up the cartridge.
With this setup, we plan to investigate:
\begin{enumerate}
    \item the selectivity of columns to different pollutants;
    \item the time of exhaustion of a column and the procedure needed to recover it.
\end{enumerate}
This investigation process will take place in the laboratory where the system will be assembled, calibrated and tested, while in a second moment a data taking campaign in the barn will then take place for a few months and data will be analyzed to understand the real efficiency of the prototype in realistic and controlled environmental and operational conditions.

\section{Evaluation of the impact of the project}\label{sec:impact}

Livestock production significantly contributes to global GHG emissions, as the Assessment Report~6 of Intergovernmental Panel on Climate Change quantified CH$_4$ livestock emissions in 123 Mt/yr, which, together with substantial N$_2$O and CO$_2$ emissions, contributed with a 12\% to global emissions~\cite{fao2}.  
Besides ruminant enteric fermentation, emissions of CH$_4$, N$_2$O, and NH$_3$ from housing and manure management are substantial, amounting in Europe to approximately 245 Mt/yr CH$_4$, 185 Mt/yr N$_2$O, and 3.6 Mt/yr NH$_3$~\cite{eurostat}.

The CH4rLie project targets the development of a scalable system for CH$_4$ and NH$_3$ capture in barns, aiming to reduce environmental impact and enable resource recovery. Though a cow emits up to 120~kg CH$_4$/yr, atmospheric dilution and ventilation limit the concentration available for capture. Based on a measured reference concentration of 20~mg/m$^3$ in a 250-animal barn, and assuming our prototype filters one-third of the air ($\sim$3000~m$^3$/day), we estimate a potential recovery of $\sim$0.9~kg CH$_4$/month, accounting for a 50\% efficiency.

The system’s viability depends on recovering trace methane efficiently. If performance is limited in barns, it could be redirected to more favorable environments like pigsties or landfills. The regeneration of the adsorbers - preferably via pressure swing rather than thermal cycles - can reduce energy use and support frequent gas extraction. 

Zeolite, the adsorbent material, is cost-effective (\$0.3–1.8/kg), with daily usage around 2.5~kg for full capture. Alongside CH$_4$, NH$_3$ recovery is foreseen as a second-phase goal for fertilizer production. Overall, the project focuses on environmental benefits, contributing to circular bioeconomy strategies and aligning with sustainable livestock and supply chain management~\cite{oliveira}.

\section{Conclusions}
Fighting CH$_4$ emissions is crucial in the climate change mitigation effort. In this context, emissions from livestock farming pose a twofold difficulty. First of all, emissions should be reduced without affecting animal life. Furthermore, capturing CH$_4$ after it has been released into the atmosphere, where its concentration is very low, is highly challenging. Hence, the CH$_4$ Livestock Emission project proposes an interesting and low-cost technological alternative that has already been validated in the very different context of the Large Hadron Collider. Conducting research in this field is essential - not only from a climate standpoint but also in terms of energy - since the captured methane could serve as a clean energy source, helping to prevent additional harmful emissions.

\section*{Author Contributions}
\begin{itemize}
    \item Conceptualization: I. Vai, S. Calzaferri, P. Salvini
    \item Data curation: I. Vai, P. Vitulo, 
    \item Formal analysis: P. Vitulo, F. A. Angiulli, N. K. Kameswaran, D. Dondi, D. Vadivel, E. Dinuccio
    \item Funding acquisition: I. Vai
    \item Investigation: F. A. Angiulli, M. C. Arena, M. Brunoldi, S. Calzaferri, E. Dinuccio, D. Dondi, N. K. Kameswaran, D. Vadivel, R. Verna
    \item Methodology: M. C. Arena, D. Biagini, E. Dinuccio, D. Dondi, R. Guida, B. Mandelli
    \item Project administration: I. Vai, E. Dinuccio, L. Finco
    \item Resources: I. Vai, E. Dinuccio, L. Finco
    \item Software: F. A. Angiulli, C. Aimè, A. Braghieri, L. Finco, C. Riccardi, P. Salvini, A. Tamigio, P. Vitulo
    \item Supervision: I. Vai
    \item Validation: F. A. Angiulli, M. C. Arena, M. Brunoldi, S. Calzaferri
    \item Visualization: I. Vai, F. A. Angiulli, M. Brunoldi, N. K. Kameswaran, E. Dinuccio
    \item Writing – original draft: I. Vai
    \item Writing – review and editing: I. Vai, P. Montagna, L. Finco, C. Aimè, S. Calzaferri, M. Brunoldi, F. A. Angiulli
\end{itemize}

\section*{Funding}
This project is funded by the European Union – Next Generation EU PRIN 2022 PNRR - P2022FTF7L.

\section*{Acknowledgments}
D. Vadivel and D. Dondi acknowledge support from the Ministry of University and Research (MUR) and the University of Pavia through the program “Dipartimenti di Eccellenza 2023–2027.”

S. Calzaferri and N. K. Kameswaran acknowledge support from the European Union – Next Generation EU PRIN 2022 PNRR - project P2022FTF7L.

\section*{Competing Interests}
The authors declare that they have no known competing financial interests or personal relationships that could have influenced the work reported in this paper.

\end{document}